\documentclass[11pt]{article}
\usepackage{epsfig}

\newcommand{\be}{\begin{equation}}
\newcommand{\ee}{\end{equation}}
\newcommand{\bea}{\begin{eqnarray}}
\newcommand{\eea}{\end{eqnarray}}
\newcommand{\beas}{\begin{eqnarray*}}
\newcommand{\eeas}{\end{eqnarray*}}
\newcommand{\bdm}{\begin{displaymath}}
\newcommand{\edm}{\end{displaymath}}
\newcommand{\ba}{\begin{array}}
\newcommand{\ea}{\end{array}}
\newcommand{\bi}{\begin{itemize}}
\newcommand{\ei}{\end{itemize}}
\newcommand{\ben}{\begin{enumerate}}
\newcommand{\een}{\end{enumerate}}
\newcommand{\bc}{\begin{center}}
\newcommand{\ec}{\end{center}}
\newcommand{\bfl}{\begin{flushleft}}
\newcommand{\efl}{\end{flushleft}}
\newcommand{\bfr}{\begin{flushright}}
\newcommand{\efr}{\end{flushright}}
\newcommand{\bmc}{\begin{multicols}}
\newcommand{\emc}{\end{multicols}}
\newcommand{\bd}{\begin{description}}
\newcommand{\ed}{\end{description}}
\newcommand{\bq}{\begin{quote}}
\newcommand{\eq}{\end{quote}}
\newcommand{\bfg}{\begin{figure}}
\newcommand{\efg}{\end{figure}}
\newcommand{\bt}{\begin{table}}
\newcommand{\et}{\end{table}}
\newcommand{\blt}{\begin{longtable}}
\newcommand{\elt}{\end{longtable}}
\newcommand{\btb}{\begin{tabular}}
\newcommand{\etb}{\end{tabular}}
\newcommand{\btg}{\begin{tabbing}}
\newcommand{\etg}{\end{tabbing}}
\newcommand{\bp}{\begin{picture}}
\newcommand{\ep}{\end{picture}}
\newcommand{\bmp}{\begin{minipage}}
\newcommand{\emp}{\end{minipage}}

\newcommand{\nop}{\rule{0cm}{0cm}}

\newcommand{\itg}{\int \limits}

\newcounter{talk}
\newcommand{\talk}[2]{\refstepcounter{talk}\newpage
           \bc {\Huge\bf\arabic{talk} \\[1cm]
           \Large#1}\\[1cm]
           #2\ec\setcounter{section}{0}%
           \setcounter{subsection}{0}\nop\\ \nop}

\newcommand{\titlestyle}{\it}
\newcommand{\volstyle}{\bf}
\newcommand{\commname}{Private communication}
\newcommand{\prepname}{(in preparation)}

\newcommand{\lessgtr}{{\raisebox{-0.4ex}[0ex][0ex]{$<$}
                       \atop \raisebox{0.3ex}[0ex][0ex]{$>$}}}

\begin{document}

\title{\nop\\[-2cm]
       \nop\hfill{\normalsize MZ-TH/96-39}\\
       \nop\hfill{\normalsize hep-ph/9611378}\\[1cm]
       \LARGE\bf Introduction to XLOOPS \\[0.5cm] \nop}

\author{L. Br\"ucher, J. Franzkowski, A. Frink, D. Kreimer \\[1cm]
\small Institut f\"ur Physik,
Johannes Gutenberg-Universit\"at Mainz \\
\small D-55099 Mainz, Germany\\[0.5cm] \nop}

\date{\small Talks presented at the \\
  Fifth International Workshop on new Computing Techniques in
  Physics Research (AIHENP 96)\\
  Lausanne, September 1996}

\maketitle

\begin{abstract}
The program package XLOOPS calculates massive one- and two-loop
Feynman diagrams. It consists of five parts: 
\begin{itemize}
 \item a graphical user interface 
 \item routines for generating diagrams from particle input 
 \item procedures for calculating one-loop integrals both analytically
   and numerically
 \item routines for massive two-loop integrals 
 \item programs for numerical integration of two-loop diagrams.
\end{itemize}
The package relies on the application of
parallel space techniques. The treatment of tensor structure and 
the separation of UV and IR divergences in analytic expressions is
described in this scheme. All analytic calculations are performed with
MAPLE. Two-loop examples taken from Standard Model calculations are
presented. The method has recently been extended to all two-loop vertex
topologies, including the crossed topology, graphs with divergent
subloops and IR divergent diagrams. This will be included in the
XLOOPS package in the near future.
\end{abstract}

\thispagestyle{empty}

\talk{XLOOPS -- an introduction \\[0.4cm]
to parallel space techniques
\protect\footnote{Work supported by grant CHRX-CT94-0579, from HUCAM.}}
{Dirk Kreimer\protect\footnote{email: 
kreimer@dipmza.physik.uni-mainz.de}}

The package {\em XLOOPS} presented in this workshop relies on the
application of parallel space techniques. We introduce these
techniques covering the following topics:
\begin{itemize}
\item    The generation of integral representations for massive
         two-loop diagrams.
\item    The treatment of tensor structures. 
\item   The handling of the $\gamma$-algebra in this scheme.
\item   The separation of UV and IR divergences in analytic expressions.
\end{itemize}
We present two-loop examples taken from Standard Model calculations.
 
\section{Introduction}

In last year's conference in Pisa, AIHENP95, David Broadhurst and myself
presented for the first time results concerning a fascinating
connection between knot theory and Feynman diagrams.$^1$ Meanwhile 
this has lead to numerous results in field theory, as well as in 
knot theory and number theory.$^2$ We are still working in that area,
and the results are even more intriguing, with the connection
between transcendental numbers arising in counterterms and knots
associated with the Feynman graphs becoming more and more specific.
David Broadhurst and I just finished a joint two month stay at the
University of Tasmania, where Bob Delbourgo invited us to continue
our collaboration on the subject. As a consequence, David had to go back
to teach at summer schools directly after our stay in Tasmania, so
that for him it is not possible to attend this conference. 

Today, I would like to focus on the techniques used in the package
{\em XLOOPS} presented at this workshop. It is based on a very simple
idea: that it might be useful to combine dimensional regularization
with the fact that each Green function offers a distinguished
subspace of spacetime - the space spanned by all its external momenta.
This very fact was used as a convenient means to define dimensional
regularization, for example in Collins text book.$^3$

Amazingly, a systematic exploration of this idea allows to make progress
for the calculation of general massive one- and two-loop Green function.
The package, as it is presented in this conference, focuses on
two- and three-point Green functions.$^4$ While Lars Br\"ucher and
Johannes Franzkowski will report on the package {\em XLOOPS} itself,
Alexander Frink will report on progress for the scalar two-loop
three-point functions, covering planar and non-planar topologies,
results which were already used in recent Standard Model
calculations.$^5$

\section{Parallel and Orthogonal Spaces}

In our approach, we use the fact that any non-trivial Green function
furnishes a set of external momenta, on which it depends.
A $n$-point Green functions provides in fact $n-1$ independent external
momenta $p_i$. This allows to write any loop momentum $k_i$
as a sum
of two covariant vectors, $k_i^\mu=k_{i,\parallel}^\mu
+k_{i,\perp}^\mu$, where $k_{i,\parallel }$ has components 
in a space which is the
linear span of the external momenta $p_i$, the parallel space, 
while $k_{i,\perp}$  is in its orthogonal complement, $\sum_\mu
k_{i,\perp}^\mu
p_{i\mu}=0$. Dimensional continuation happens in the orthogonal space.

\subsection{Two-point functions}

To consider the first specific example, we turn to a massive one-loop
two-point function, providing a scalar integral of the form
\begin{equation}
I_2:=\int d^D\!k\;\frac{1}{[k^2-m_1^2][(k+q)^2-m_2^2]}\equiv\int
\frac{d^D\!k}{P_1\,P_2}.
\end{equation}
In parallel and orthogonal space variables, we have
\begin{eqnarray}
I_2 & = & \frac{2\pi^{\frac{D-1}{2}}}{\Gamma(\frac{D-1}{2})}
\int_{-\infty}^\infty dk_\parallel \int_0^\infty dk_\perp
\frac{k_\perp^{D-2}}{[k_\parallel ^2-k_\perp^2-m_1^2][(k_\parallel +q)^2-k_\perp^2-m_2^2]}\nonumber\\
 & = & \frac{2\pi^{\frac{D-1}{2}}}{\Gamma(\frac{D-1}{2})}
\int_{-\infty}^\infty dk_\parallel  \frac{1}{P_2-P_1}
\int_0^\infty dk_\perp \left[ 
\frac{k_\perp^{D-2}}{P_1}-\frac{k_\perp^{D-2}}{P_2}
\right],
\end{eqnarray}
where
\begin{equation}
k^\mu=k_\parallel e_{\parallel}^\mu+k_{\perp}^\mu,\,\,
e_{\parallel}^\mu=\frac{q^\mu}{\sqrt{q^2}},\,\,
k_\perp\cdot q=0. 
\end{equation}
We straightforwardly integrate the orthogonal space to find expressions
which involve
\begin{equation}
\left[ P_i\mid_{k_\perp=0}\right]^{\frac{D-3}{2}},
\end{equation}
which determine the cut structure of the result. Note that we
avoid the Wick rotation and Feynman parametrizations altogether.
The remaining integration with respect to $dk_\parallel $ can be done easily.
For example, we can reinterprete it as an integral representation of a
${\cal R}$-function.$^6$ Upon expanding in $D-4$, we recover the standard
results of perturbation theory.

If we go to the two-loop level, we have five non-trivial integrations.
We had two for the one-loop case, and for the two-loop case we have one
more than twice this number, as there is an extra angular integration
in orthogonal space, due to the presence of a non-trivial scalar 
product $l_\perp \cdot k_\perp$.
It is known that straightforward application of the parallel space
method gives a two-fold integral representation.$^7$
This integral representation is the basis for the two-loop two-point
package included in $XLOOPS$.

\subsection{Three-point functions}

Now we have a two-dimensional parallel space. Accordingly, in the
one-loop case, the scalar one-loop integral has the following form
in parallel and orthogonal space variables
\begin{eqnarray}
I_3 & = & \frac{2\pi^{\frac{D-2}{2}}}{\Gamma(\frac{D-2}{2})}
\int_{-\infty}^\infty d\!k_0 d\!k_1 \frac{1}{[P_i-P_j][P_i-P_k]}
\int_0^\infty d\!k_\perp \frac{k_\perp^{D-3}}{P_i}
+\mbox{perm.~in $i,j,k$},\nonumber\\
P_1 & = & k_0^2-k_1^2-k_\perp^2-m_1^2,\nonumber\\
P_2 & = & (k_0+q_{1,0})^2-(k_1+q_{1,1})^2-k_\perp^2-m_2^2,\nonumber\\
P_3 & = & (k_0+q_{2,0})^2-(k_1+q_{2,1})^2-k_\perp^2-m_3^2,
\end{eqnarray}
where
\begin{eqnarray}
k_{\parallel}^\mu=k_0 e_{1}^\mu+k_1 e_{2}^\mu,\,\,
k^\mu=k_{\parallel}^\mu+k_{\perp}^\mu,\nonumber\\
e_1\cdot k_\perp=e_2\cdot k_\perp=0,\, span\{e_1,e_2\}=span\{q_1,q_2\}.
\end{eqnarray}
There are two non-trivial integrations
for the parallel space, assumed to be spanned by two external
momenta $q_1,q_2$, and one for the modulus of $k_\perp$.
One can proceed by explicitly using the signature of spacetime in the
parallel space: a shift $k_0 \to k_0+k_1$ 
renders all propagators linear in the
variable $k_1$. The orthogonal space integration confronts us
with a non-trivial cut structure: the integrand obtains a factor 
$[P_i\mid_{k_\perp=0}]^{\frac{D-4}{2}}$. 
Due to the shift in $k_0$, the cut is confined to a half
plane in complex $k_1$ space, so that we can use the residue theorem for
one of the parallel space integrations. The resulting expression
can be identified as an integral representation of a $R$-function or,
equivalently, of a hypergeometric.$^6$

In the two-loop case, some of the features remain. We are now confronted
with four integrations for the parallel space, and three for the
orthogonal space. As in the two-point case, we do the angular
integration in the orthogonal space, with respect to
$z=\frac{l_\perp\cdot k_\perp}{\mid l_\perp\mid\!\mid k_\perp\mid}$, 
first. This results in a non-trivial
cut structure, and the remaining integrations are determined by the
demand that as many as possible can be done as residue integrals
by closing the contour at infinity. A residue contributes only if it is
located in the interior of the contour, and this results in constraints
for the remaining variables. It turns out that these constraints
determine
the remaining domains of integrations to be finite triangles
in the space of two remaining
parallel space variables. 
At this level we are talking about a four-fold integral
representation. Two further integrations in the orthogonal space
variables $l_\perp^2,k_\perp^2$ can be achieved for the planar topology,
resulting in a two-fold integral representation, 
by using Euler transformations.$^8$
Alexander Frink achieved a similar result for the non-planar topologies,
which was already used and tested recently.$^5$ He will report
on these achievements in his talk in detail.
 
\subsection{The general case}

One could extend the method to four-point Green functions
(and higher). We have not included this at this stage, but do not see
any conceptual problems to do so if necessary.

\section{Tensor Structures and $\gamma$-Algebra}

One of the most interesting 
features of our approach is its treatment of the
spin structure of Green functions. 
We altogether avoid the calculation of single tensor integrals
(though this is possible in {\em XLOOPS}, to make contact with more
conventional methods), but 
directly calculate the characteristic polynomials of a graph.$^9$ 

\subsection{Two-point functions}

Again let us go back to a one-loop two-point function to understand the
idea. Any Green function will deliver polynomial expressions
in $k_0,\,k_\perp^2$ for its numerator. They might come distributed
over various form factors.$^9$ 
To handle this situation, we
also separate the Clifford algebra into two orthogonal sub-algebras,
according to the splitting into parallel and orthogonal spaces. 
Using basic properties of Clifford algebras, it is then easy to
determine the characteristic polynomials. 
A trivial one-loop example is provided by the self energy
of a fermion, dressed with a massless vector boson in the Feynman gauge
\begin{eqnarray}
\int dk_\parallel \,dk_\perp \frac{\gamma_\mu [k_\parallel \gamma_\parallel -\gamma_\perp\cdot
k_\perp+m]\gamma^\mu}{[(k+q)^2][k^2-m^2]}\rightarrow
\int dk_\parallel \,dk_\perp \frac{[(2-D)k_\parallel \gamma_\parallel +m{\bf 1}]}{[(k+q)^2]
[k^2-m^2]}.
\end{eqnarray}
We have two formfactors, one proportional to $\gamma_\parallel $, 
the other
one proportional to ${\bf 1}$ in spin space. In parallel
space, the only element of the $\gamma$-algebra is 
$\gamma_\parallel ^\nu = q^\nu q^\mu\gamma_\mu/q^2=\gamma\cdot e$, 
while the 
Clifford algebra in orthogonal space is spanned by elements
$\gamma_\perp^\nu$, with $q\cdot\gamma_\perp=0$ 
and $\{\gamma_\parallel ,\gamma_\perp\}=0$.$^9$
In the above example we found two very simple
characteristic polynomials for the two formfactors, $(2-D)k_\parallel $
and $m$.

Note that any polynomial expression in $k_\parallel ,\,k_\perp^2$ can be
calculated by either reducing the expression to some basic
scalar integrals, or to trivial massive tadpoles, by using the following
rules
\begin{eqnarray}
k_\parallel  & = & \frac{P_2-P_1+[m_2^2-m_1^2-q^2]}{2\sqrt{q^2}},\nonumber\\
k_\perp^2 & = & k_\parallel ^2-P_1-m_1^2,\nonumber\\
\int d^Dk \frac{[k_\perp^2]^r}{P_1\,P_2} & = & \int d^{D+2r}\frac{1}{P_1\,P_2}.
\end{eqnarray}

Johannes Franzkowski
will have further comments on the implementation of these ideas
at the two loop level in his talk.  

\subsection{Three-point functions}

Here we have the possibility to reduce polynomial expressions in the 
variables $k_0,\,k_1,\,k_\perp^2$ to either two-point functions, 
or to basic scalar three-point
functions. Again, we achieve this by solving for the above
variables in terms of propagators, masses and external momenta.
These reductions are already implemented in {\em XLOOPS} in the
one-loop case, and will be available at the two-loop case in the near
future.

\section{UV and IR Divergences}

We usually do a termwise determination of the UV degree of divergence
for our characteristic polynomials. We subtract massless integrals
to achieve finite integral representations, carefully avoiding
oversubtractions and thus avoiding the generation of spurious
infrared singularities.$^9$ A main feature of our approach is 
that the subtraction is 
done termwise in the characteristic polynomial. This avoids
oversubtractions for the non-leading terms in the characteristic
polynomials, while only the powercounting for the leading terms in the 
characteristic
polynomial agrees with the powercounting for the graph itself.

At the moment, we gain first experience for the subtraction of IR
singularities from our integral representations in collaboration with
Jochem Fleischer in Bielefeld.

At the two-loop level, we calculate the UV divergences of a Feynman
graph analytically, and generate well-defined code for numerical
integrations.$^9$ 

\section{Conclusions}

Let me list our main results as follows:
\begin{itemize}
\item {\em XLOOPS} calculates arbitrary graphs in the Standard Model
for one-loop two- and three-point functions. After specifying a consistent
particle content for a chosen topology, the user can choose to get
an analytic or numeric result. 
\item At the two-loop level, {\em XLOOPS} returns an analytic result
for the UV-divergent part of any two-point Standard Model graph,
and automatically generates code for the remaining finite part
of the whole graph.
\item We have integral representations
for all scalar two-loop three-point functions and will implement 
the two-loop three-point case in {\em XLOOPS} in the future. 
\end{itemize}

\section*{Acknowledgements}

Foremost, I like to thank Lars Br\"ucher, Johannes Franzkowski and
Alexander Frink for the patience with which they transformed
an idea into concrete tools for two-loop calculations.
I also like to thank J\"urgen K\"orner and Karl Schilcher
for stimulating discussions. It is a pleasure to thankBob Delbourgo for
hospitality at the University of Tasmania.

\talk{XLOOPS -- a package calculating \\[0.4cm]
one- and two-loop diagrams}
{Lars Br\"ucher\protect\footnote{e-mail:
bruecher@dipmza.physik.uni-mainz.de}}

 A program package for calculating massive one- and two-loop
  diagrams is introduced. It consists of five parts: 
\begin{itemize}
 \item a graphical user interface 
 \item routines for generating diagrams from particle input 
 \item procedures for calculating one-loop integrals both analytically
   and numerically
 \item routines for massive two-loop integrals 
 \item  programs for numerical integration of two-loop diagrams.
\end{itemize}
  Here the graphical user interface and the text interface to Maple
  are presented.
 
\section{Introduction}    

It is a well known fact, that high precision calculations of reactions
between elementary particles cannot be done without the aid of
computer programs, as the number of Feynman graphs can easily reach
one hundred and exceed this as well. So there have been developed several
different packages, each solving only one aspect or part of such
calculations. More amazing is the fact that there is no package
covering the whole procedure of calculating graphs. So someone
starting such a high precision calculation is confronted with the
fact, that he has to learn the syntax of several different programs.
To overcome this problem, the development of XLOOPS, a program
package covering all aspects of the calculation of Feynman graphs up
to two-loop level was started.  
The following sections will discuss the structure of XLOOPS and
introduce the graphical user interface
(GUI), which makes XLOOPS an `easy-to-handle' program package.  
 
\section{The general structure of XLOOPS}

To avoid being tied to special computer systems or architectures,
XLOOPS is designed to be as much portable as possible.
Therefore 
\begin{itemize}
 \item Maple V\cite{Ma1,Ma2},
 \item TCL/Tk\cite{TCL}and
 \item C++\cite{CC} for numerical integrations
\end{itemize}
were chosen as programming languages, as they are available for 
almost every computer architecture and all common operating systems,
like Unix, Windows (3.1, 95 and NT) and  VMS.
\begin{figure}[htbp]
\epsfig{file=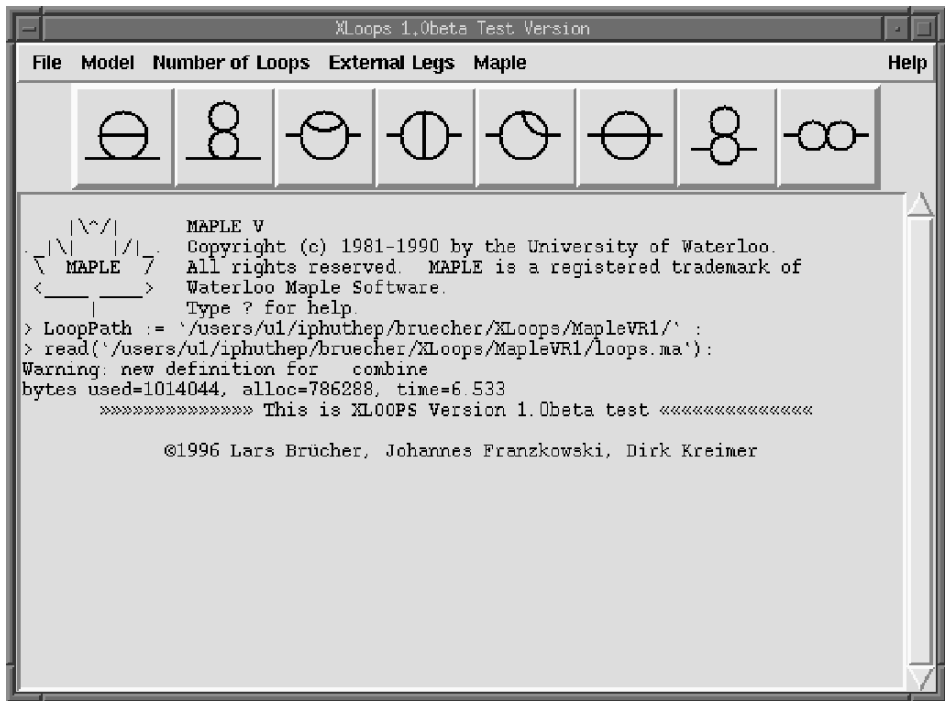,width=12.5cm}
\caption{The main window of XLOOPS, which appears after starting the program}
\label{pic1}
\end{figure}

This choice of languages can also be seen as an indicator for the
program structure of XLOOPS.
As already mentioned in the abstract XLOOPS consist of
\begin{itemize}
 \item a graphical user interface 
 \item routines for generating diagrams from particle input 
 \item procedures for calculating one-loop integrals both analytically
   and numerically
 \item routines for massive two-loop integrals 
 \item  programs for numerical integration of two-loop diagrams.
\end{itemize}
The graphical user interface should make the input of Feynman graphs as convenient as
possible. This part is written in TCL/Tk and includes procedures for
translating the user's input into commands suitable for the
following parts. These parts, the second, third and fourth part, are
written in Maple V. They provide procedures which insert
Feynman rules and evaluate loop-integrals analytically as far as
possible. The user might also use these procedures without the
graphical front end. The fifth part is the numerical integration
program for the remaining integrals in two-loop calculations which
is automatically invoked if needed. This part will be covered by the
talks of J. Franzkowski and A. Frink.

When invoking XLOOPS the main window appears (see Fig. \ref{pic1}). It
provides an overview of all topologies for a $n$-point $m$-loop
function with $n$ and $m$ defined
by the user in the main menu. By clicking on a topology the  
chosen graph appears in an extra window, suitable for inserting 
all virtual and non-virtual particles reacting (see Fig. \ref{pic2}).
Having inserted the
particles one can choose whether the program should evaluate the result
expressed in loop-integrals by clicking on the `Evaluate' button or if
the program should return also the loop-integral evaluated in
logarithms and dilogarithm by clicking on the `Evaluate Full' button.
The program itself starts now the calculation of the graph by giving
an appropriate command to the Maple V part of the package, which
actually gives back the analytical result displayed in the Maple
output section of the main window. If a two-loop calculation is
performed and all numerical values are given, the program also starts
the numerical integration of the remaining integrals.

The result obtained by the evaluation can be saved or written out as a C
program to a file. This is convenient for later use in a plotting
program for example. For additional symbolic manipulations of the result,
like inserting the renormalization conditions, a special
window for direct Maple input exists.
Moreover the program provides as additional features the possibility to
insert the latest values from Particle Data Group\cite{DATA} for the particle
properties as a menu entry. 

\section{Examples calculated with XLOOPS}

\subsection{The decay $\bf H^+ \longrightarrow W^+ h^0$}

To get more specific, the evaluation of $H^+ \longrightarrow W^+ h^0$
in the framework of the Two Higgs Doubletts Model\cite{rui1} will be illustrated
in this section. As an example we choose the top loop contribution to the
self energy $H^0
\rightarrow h^0$ necessary for the wave function renormalization and
the triangle graph, which represents the actual one-loop
correction (for a detailed discussion see \cite{1}).

\begin{figure}[htbp]
\epsfig{file=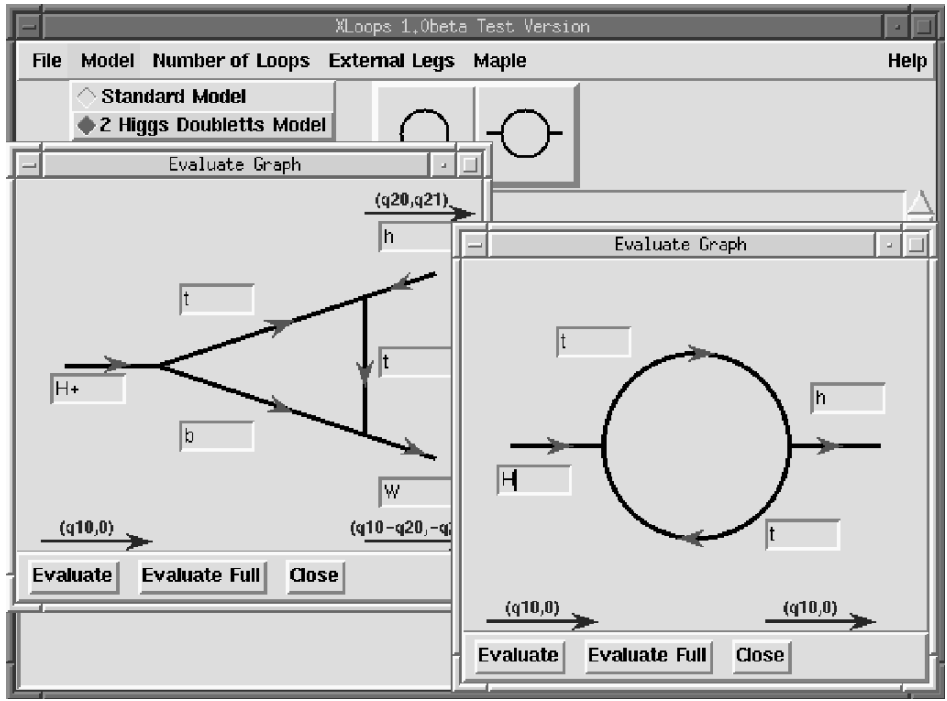,width=12.5cm}
\caption{non-diagonal self energy graph and triangle graph
  contributing to $H^+ \rightarrow W^+ h^0$ }
\label{pic2}
\end{figure}

After choosing the `2 Higgs Doubletts Model' from the main menu and
clicking on the graph we want to calculate, the window where the
particles have to be inserted appears (see Fig. \ref{pic2}). When
finished with typing in the particles and clicking on `Evaluate Full'
the following result for the $H^0\rightarrow h^0$ graph appears in the Maple
output section:

{\footnotesize
\begin{verbatim}
G0 := [
            2     2                             2         2
           e  Mtop  sin(alpha) cos(alpha) I (q10  - 4 Mtop )
C1 = [1/32 -------------------------------------------------,
                             2   2          2   2
                      sin(tw)  Mw  sin(beta)  Pi

      2     2                                 2   2      2     2         2
1/64 e  Mtop  sin(alpha) cos(alpha) (2 Ln(4 Pi  MU ) I Pi  (q10  - 4 Mtop ) - 2

     2           2           2         2            2        2
   Pi  (- 2 I q10  + 6 I Mtop  + Pi q10  - 2 Pi Mtop  + I q10  gamma

          2                  2                 2                2        2
   + I q10  Ln(Pi) - 4 I Mtop  gamma - 4 I Mtop  Ln(Pi) - I Mtop  Ln(Mtop )

           2        2
   - I Mtop  Ln(Mtop  - I rho)

                                                     2                     2
   - I (%1 (Ln(1 - %1) - Ln(1 + %1) + I Pi) - Ln(Mtop  - I rho) - I Pi) q10

                                                       2                      2
   + 2 I (%1 (Ln(1 - %1) - Ln(1 + %1) + I Pi) - Ln(Mtop  - I rho) - I Pi) Mtop

       /         2   2          2   4
  ))  /  (sin(tw)  Mw  sin(beta)  Pi )
     /

],     C1]
                                          2
                                    - Mtop  + I rho
%1 :=                    Sqrt(1 + 4 ---------------)
                                             2
                                          q10

\end{verbatim}
}

After evaluating the tadpole contribution in a similar way one can
insert the on-shell condition, which in this case reads:
\begin{equation}
  \delta Z_{H^0 h^0}^{(1)} = 
  \frac{1}{m_{h^0}^2-m_{H^0}^2} \Sigma_{H^0 h^0}(m_{h^0}^2)
\end{equation}
This condition can be directly passed to Maple.
In a similar manner the final result, after having
summed up all graphs, can also be exported as C language code. This is
very helpful for producing plots as shown in \cite{1}.

\subsection{Flavour changing self-energy}

As a second example the flavour changing self energy\cite{self} from $s
\rightarrow d$ in the
framework of the Standard Model will be shown. 

\begin{figure}[htbp]
\epsfig{file=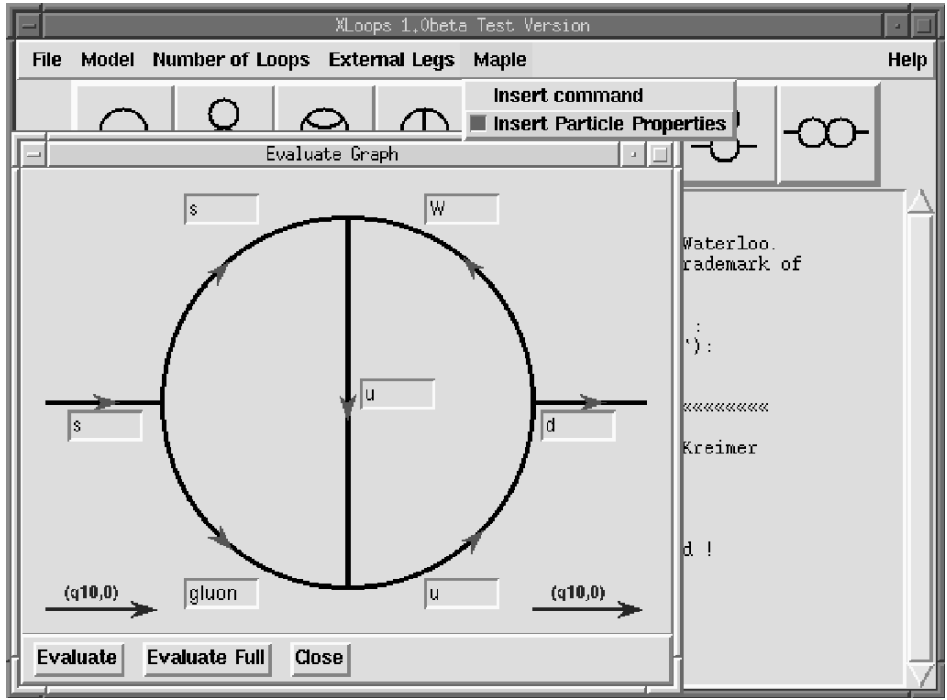,width=12.5cm}
\caption{The self energy }
\label{pic4}
\end{figure}

After having inserted the particle properties from the main menu entry
and having attached the appropriate particle at each particle line as
shown in Fig. \ref{pic4} the evaluation is again started with
`Evaluate Full'. In this case the program first starts the Maple part
to evaluate the divergent part of the integral and the two-fold integral
representation suitable for later numerical integration. As all
numerical values are given, XLOOPS starts the remaining numerical
integration. The result is displayed in the main window as a function of
the form factors. Each of the form factors' coefficients is displayed
as a series in $D-4$.
   
\section{Availability and Outlook}

Currently a demo version of the program package, which does not
include two-loop calculations, is available at 
\begin{displaymath}
  http://dipmza.physik.uni-mainz.de/\sim bruecher/xloops.html
\end{displaymath}
where also additional information is accessible. The full version,
which currently evaluates all one-loop graphs up to the 3-point
function and all two-loop graphs up to the 2-point function, will
also be accessible there, when it is fully tested. At the moment work
to incorporate the 2-loop 3-point and the 1-loop 4-point function is
in progress.

In future, procedures that automatically draw all Feynman graphs for
a given process and display them in Postscript format should be
added as well as procedures for automatic renormalization.

\section*{Acknowledgements}

I would like to thank K. Schilcher, J. K\"orner, D. Kreimer and
J.B. Tausk for many
helpful discussions on this topic as well as J. Franzkowski and
A. Frink for not desperating while testing beta software. This work is
supported by the `Graduiertenkolleg Elementarteilchenphysik bei hohen
und mittleren Energien' from University of Mainz.

\talk{Automatic calculation of massive \\[0.4cm]
two-loop self-energies with XLOOPS}
{Johannes Franzkowski\protect\footnote{e-mail:
franzkowski@dipmza.physik.uni-mainz.de}}

Within the program package XLOOPS it is possible to calculate
self-energies up to the two-loop level for arbitrary massive particles. The
program package -- written in MAPLE \cite{MA1,MA2} -- is designed to deal with
the full tensor structure of the occurring integrals. This means that
applications are not restricted to those cases where the reduction to scalars
via equivalence theorem is allowed. 

The algorithms handle two-loop integrals analytically if this is possible.
For those topologies where no analytic result for the general mass case is
available, the diagrams are reduced to integral representations which
encounter at most a two-fold integration. These integral representations are
numerically stable and can be performed easily using VEGAS \cite{Ve1,Ve2}.

\section{Structure of the XLOOPS package}

The aim of XLOOPS is to provide the user with a program package for complete
evaluation of Feynman diagrams. The package consists of the following parts:
\bi
\item Input via Xwindows interface: \\
  In a window the user selects the topology which shall be calculated. A
  Feynman diagram pops up in which the particle names have to be inserted.
\item Processing with MAPLE: \\
  The selected diagram is evaluated. The necessary steps for reducing the
  numerator -- the so-called characteristic polynomial -- are performed by
  routines written for MAPLE. The result is expressed in terms of one- and
  two-loop integrals.
\item Evaluation of one-loop integrals: \\
  One-loop one-, two- and three-point integrals are calculated analytically or
  numerically to any tensor degree using MAPLE.
\item Evaluation of two-loop integrals: \\
  All two-loop two-point topologies including tensor integrals are supported
  by the MAPLE routines. For those topologies where no analytic result is known,
  XLOOPS creates either an analytic two-fold integral representation or
  integrates numerically with the help of VEGAS using C++. The other topologies
  can be calculated analytically or numerically like the one-loop integrals.
\ei
In the remainder of this report the MAPLE routines, especially the treatment of
two-loop integrals is described in detail.

\section{The MAPLE part}

\subsection{Input}

The user's input is inserted conveniently by using the Xwindows interface --
there is another contribution by L. Br\"ucher which describes this interface in
more detail -- but it can also be typed in directly in a MAPLE session if the
conventions of the manual \cite{Fr7} are respected.

In any case the XLOOPS routines expect as input either a Feynman diagram or a
single integral. In XLOOPS code Feynman diagrams look like \\[0.4cm]
\hspace*{1cm}\verb+EvalGraph1(3,[bottom,charmbar,wp,bottombar,gluon,+ \\
\hspace*{3.6cm}\verb+gluon,charm,charmbar,bottom]);+ \\
\hspace*{1cm}\verb+EvalGraph2(6,[up,upbar,gluon,upbar,up,gluon,+ \\
\hspace*{3.6cm}\verb+gluon,up,upbar,up,upbar,gluon]);+ \\[0.4cm]
\verb+EvalGraph1+ calculates one-loop, \verb+EvalGraph2+ two-loop diagrams.
The first argument denotes the number of the topology, the second is the list of
particles. Single integrals have the following notation: \\[0.4cm]
\hspace*{1cm}\verb+OneLoop3Pt(+$p_0,p_1,p_2$\verb+,q10,q20,q21,m1,m2,m3);+ \\
\hspace*{1cm}\verb+TwoLoop2Pt2(+$p_0,p_1,r_0,r_1,s$\verb+,q10,m1,m2,m3,m4,m5);+
\\[0.4cm]
The $p_i,r_i,s$ determine the tensor structure $l_0^{p_0}\,l_1^{p_1}\,
l_\perp^{p_2}$ and $l_0^{p_0}\,l_\perp^{p_1}\,k_0^{r_0}\,k_\perp^{r_1}\,z^s$
respectively -- in the notation of section 2.3. The other arguments
describe momentum components and masses. A detailed list of all conventions will
be given in the XLOOPS manual \cite{Fr7}, the one-loop sector is also described
separately \cite{Fr4}.

\subsection{Characteristic polynomial}

In the case of a single integral XLOOPS skips the following steps. If a complete
diagram has to be evaluated, the routines proceed as follows:
\bi
\item The topology -- that means the information how the internal lines are
  connected -- is generated.
\item The Feynman rules are inserted. At present XLOOPS knows all Feynman
  rules of the Standard model (electroweak and QCD) and its extension to two
  Higgs doublets. The incorporation of additional models is simple.
\item The symmetry factor of the diagram is determined.
\item The characteristic polynomial (the numerator of the diagram) is evaluated
  in terms of parallel and orthogonal space variables. For that purpose XLOOPS
  knows the rules for the SU(N) algebra and how to reduce strings of Dirac
  matrices. If necessary it takes the trace of the Dirac matrices.
\item Finally the code is expressed in terms of one- and two-loop integrals.
\ei

\subsection{Parallel and orthogonal space}

The evaluation of the Dirac algebra as well as of the integrals simplifies if
one makes use of the splitting of the momentum components in parallel and
orthogonal space variables -- see D. Kreimer's contribution for details. The
definition is simple: The parallel space describes the sub-space spanned by the
external momenta, whereas the orthogonal space is the orthogonal complement of
the parallel space.

In the two-point case only one external momentum is present. Therefore the
parallel space is one-dimensional. The loop momenta $l$ and $k$ are written as
\beas
l_{0} & = & \frac{l\cdot q}{\sqrt{q^2}} \qquad 
\mbox{(projection in direction of $q$)} \\
l_{\bot} & = & \sqrt{l_{0}^2 - l^2} \qquad 
\mbox{(the same decomposition holds for $k$)} \\
l^{2} & = & l_{0}^{2} - l_{\bot}^{2} \\
k^{2} & = & k_{0}^{2} - k_{\bot}^{2} \\
l\cdot k & = & l_{0} k_{0} - l_{\bot} k_{\bot} \cos \vartheta
\eeas
The integration measure simplifies in the following sense:
\bi
\item one-loop case:
  \bdm
  \int \! d^{D}l = \itg_{-\infty}^{\infty} \!\! dl_{0} \int \!
  d^{D-1} l_{\bot} = \frac{2\pi^{\frac{D-1}{2}}}{\Gamma(\frac{D-1}{2})} \,
  \itg_{-\infty}^{\infty} \!\! dl_{0} \itg_{0}^{\infty} dl_{\bot} \, 
  l_{\bot}^{D-2} 
  \edm
  The $D$-dimensional integral reduces to two one-dimensional integrations.
\item two-loop case:
  \beas
  \lefteqn{\int \! d^{D}l \, \int \! d^{D}k = \itg_{-\infty}^{\infty} \!\! 
  dl_{0} \itg_{-\infty}^{\infty} \!\! dk_{0} \, \int \! d^{D-1} 
  l_{\bot} \int \! d^{D-1} k_{\bot}} \\
  & = & \hspace*{-0.2cm}\frac{4\pi^{\frac{D-1}{2}}\pi^{\frac{D-2}
  {2}}}{\Gamma(\frac{D-1}{2})\Gamma(\frac{D-2}{2})} \itg_{-\infty}^{\infty} \!\!
  dl_{0} \itg_{-\infty}^{\infty} \!\! dk_{0} \, \itg_{0}^{\infty} dl_{\bot} 
  \, l_{\bot}^{D-2} \itg_{0}^{\infty} dk_{\bot} \, k_{\bot}^{D-2} 
  \itg_{0}^{\pi} d\vartheta \, \sin^{D-3} \vartheta
  \eeas
  The $2D$-dimensional integral is now replaced by five one-dimensional
  integrations.
\ei

\section{Two-loop integrals}

The aim of our two-loop routines is not only to solve special mass cases or
kinematical regions. They also supply the general case where all internal masses
are different and the external momenta can be completely arbitrary. Full tensor
structure and not only the scalar case is supported. As a consequence these
general routines cannot return analytic results for all topologies. Therefore we
adopt the following strategy:
\ben
\item separate the divergent parts (in any case analytically calculable)
\item find a two-fold integral representation (in $D=4$)
\item integrate numerically the two-fold integral
\een
In the following we comment these three items.

\subsection{UV divergent integrals}

For the separation of divergences it is necessary to find a convenient
subtraction term. If a two-loop integral is multiplied by a term like the one in
brackets
\beas
\lefteqn{\int d^D l d^D k \,\,\, \frac{l_{\mu_1}\cdots l_{\mu_n}}{{\cal
P}_1(l){\cal P}_2(l){\cal P}_3(l+k){\cal P}_4(k){\cal P}_5(k)}} \\
& \to & \int d^D l d^D k \,\,\, \frac{l_{\mu_1}\cdots 
l_{\mu_n}}{{\cal P}_1(l){\cal P}_2(l){\cal P}_3(l+k){\cal P}_4(k){\cal P}_5(k)}
\,\,\left(1-\frac{{\cal P}_1(l){\cal P}_2(l){\cal P}_3(l+k)}
{l^{4}(l+k)^{2}}\right)^m
\eeas
it can be shown that the degree of divergence is decreased by $m$
steps \cite{Kr6}. To be specific, in the case of a logarithmic divergent
integral which corresponds to the case $m=1$ one gets a difference of two terms
which turns out to be convergent. To recover the original integral it is
necessary to add the subtraction term again.
\beas
& & \hspace*{-1cm}\underbrace{\int d^D l d^D k \,\,\, \frac{l_\mu l_\nu}
{{\cal P}_1(l){\cal P}_2(l){\cal P}_3(l+k){\cal P}_4(k){\cal P}_5(k)} - \int d^D
l d^D k \,\,\, \frac{l_\mu l_\nu}{l^{4}(l+k)^{2}{\cal P}_4(k)
{\cal P}_5(k)}}_{\mathrm{convergent}} \\
& & + \underbrace{\int d^D l d^D k \,\,\, \frac{l_\mu l_\nu}
{l^{4}(l+k)^{2}{\cal P}_4(k){\cal P}_5(k)}}_{\mathrm{divergent}}
\eeas
The first line which denotes the convergent part is treated in $D=4$ and
will be reduced to a two-fold integral representation which is solved
numerically. The second line contains the divergent part which can be
solved analytically in $D\not=4$.

\subsection{Integration strategy}

We integrate directly in momentum space. With the splitting in parallel and
orthogonal space variables one gets for two-point functions:
\beas
\int d^4l \int d^4k = 8\pi^2\!\!\!\!\underbrace{\itg_{-\infty}
^\infty\!\!\!dl_0dk_0}_{\mathrm{numerically}} \underbrace{\itg_0^\infty
\!\!l_\perp^2dl_\perp k_\perp^2dk_\perp \itg_0^\pi\!\!\sin\vartheta d\vartheta}
_{\mathrm{analytically}}
\eeas
The $l_0$ and $k_0$ integrations are left for numerical evaluation. All other
integrations are performed analytically.

Up to now we solved all two-point topologies for scalar and tensor evaluations.
In the three-point case all topologies are solved for the scalar case. This will
be subject of the contribution by A. Frink. At present our methods are
applied to the three-point tensor and the four-point scalar integrals.

In the following we concentrate on the two-point topologies: \\[0.5cm]
\parbox[t]{3.5cm}
{\epsfig{file=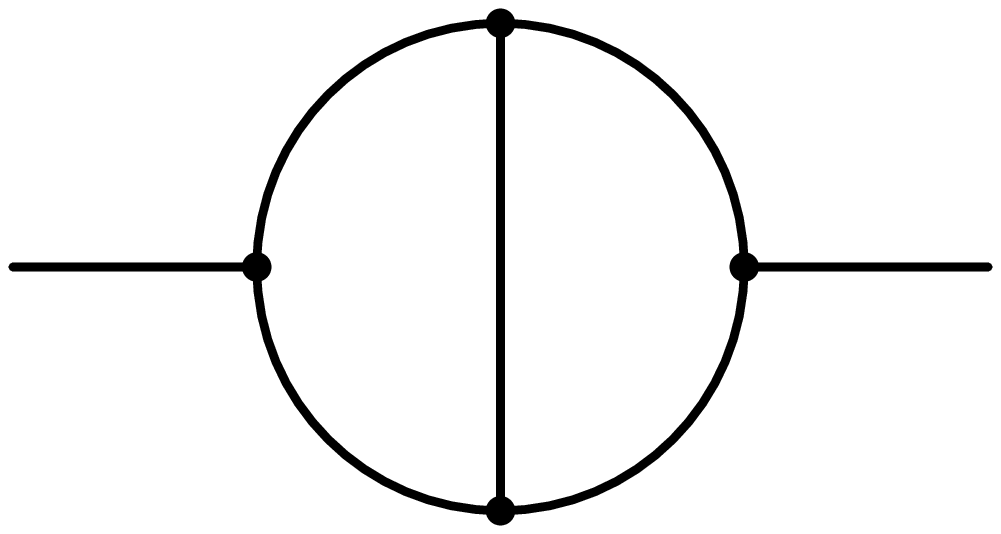,width=3cm} \\[0.4cm]
\epsfig{file=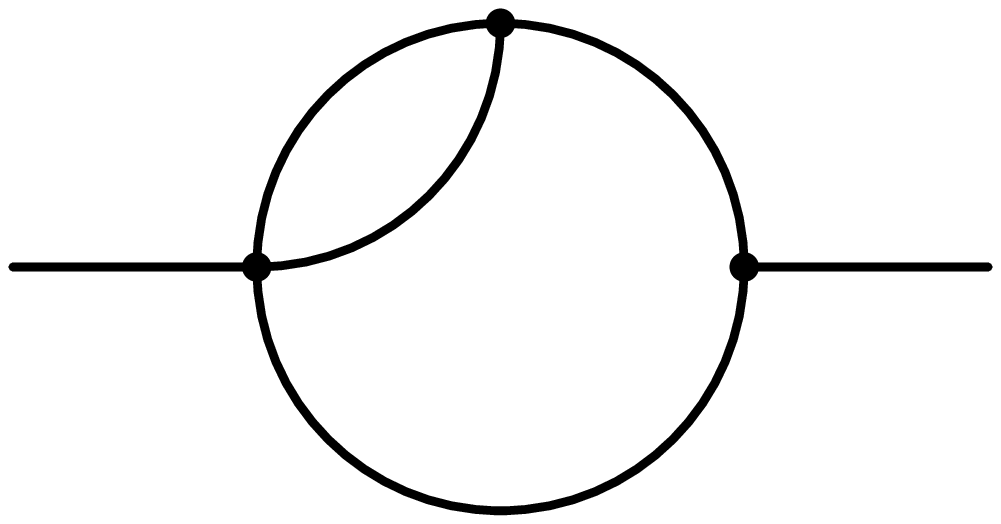,width=3cm}} \hfill
\parbox[t]{3.5cm}
{\epsfig{file=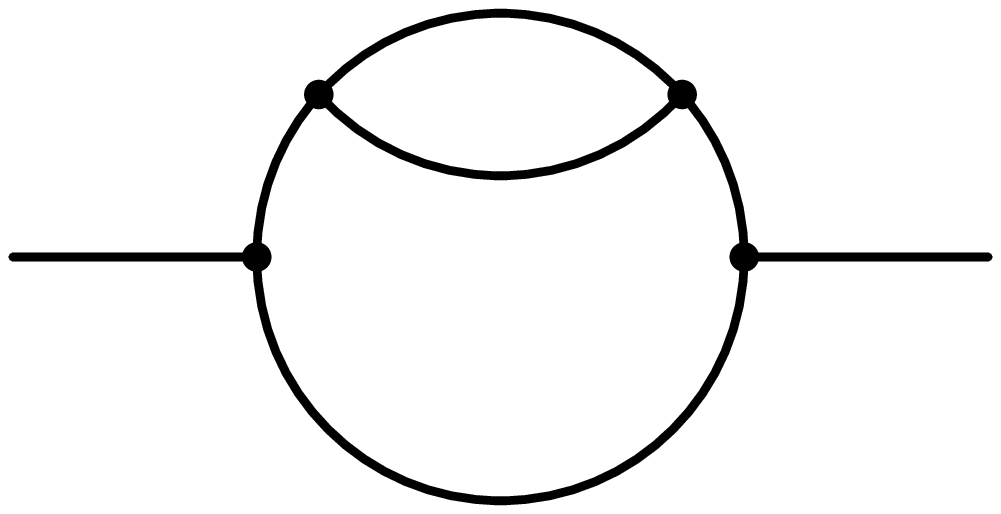,width=3cm} \\[0.8cm]
\epsfig{file=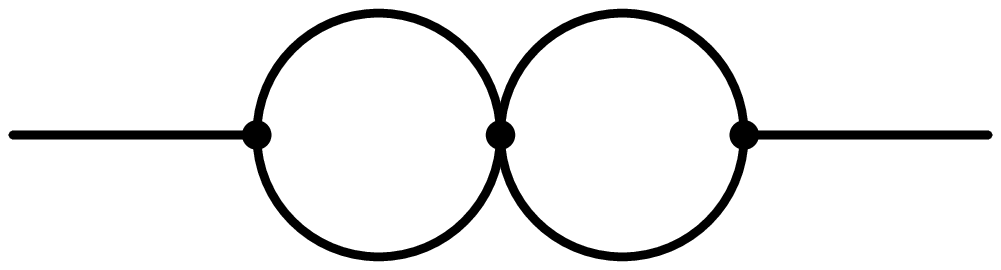,width=3cm}} \hfill
\parbox[t]{3.5cm}
{\epsfig{file=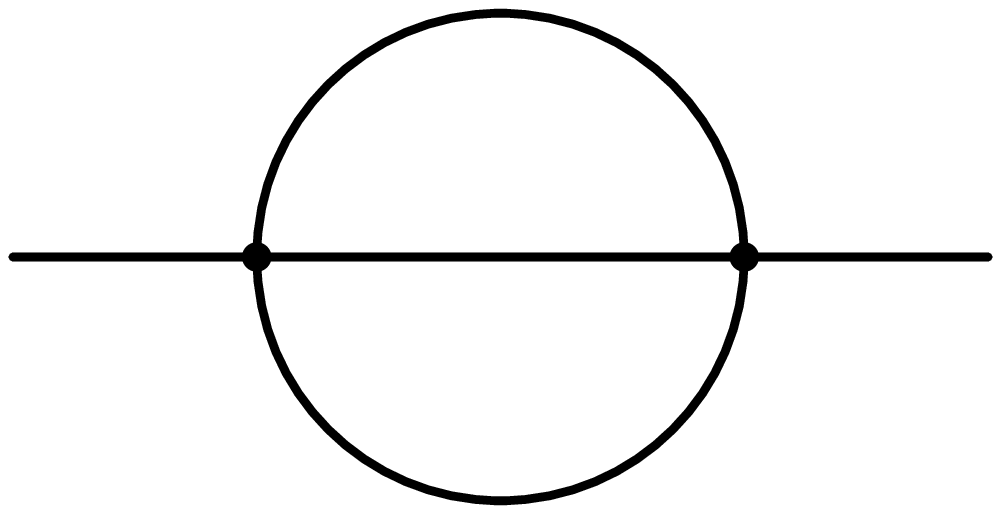,width=3cm} \\[0.4cm]
\epsfig{file=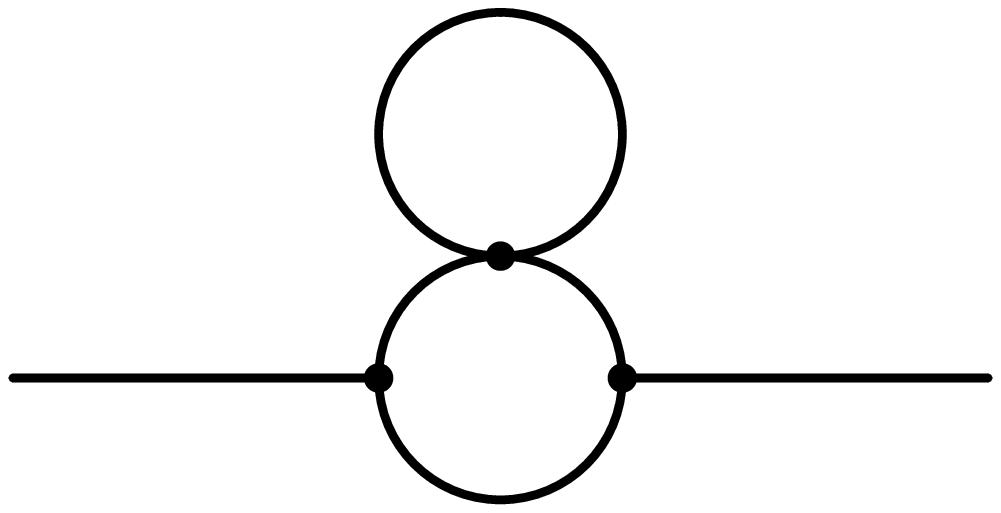,width=3cm}} \\[0.4cm]
The worst case is of course the master topology \cite{Kr4}:
\beas
\lefteqn{\int d^4l\int d^4k \,\,\frac{l_0^a\,l_\perp^b\,k_0^c\,k_\perp^d}
{{\cal P}_1 \cdots {\cal P}_5}} \\
& = & 8\pi^2\itg_{-\infty}^{\infty} dl_{0}\itg_{-\infty}
^{\infty}d k_{0}\itg_0^{\infty} d l_{\bot}\itg_0^{\infty} d k_{\bot}
\itg_{0}^{\pi} d \vartheta\,\,\,\frac{l_0^a\,l_{\bot}^{b+2}\,k_0^c\,
k_{\bot}^{d+2}\sin\vartheta}{{\cal P}_1 \cdots {\cal P}_5} \\[0.2cm]
{\cal P}_1&=&l_{0}^2-l_{\bot}^2-m_1^2+i\varrho \\
{\cal P}_2&=&(l_{0}+q)^2-l_{\bot}^2-m_2^2+i\varrho \\
{\cal P}_3&=&(l_{0}+k_{0})^2-l_{\bot}^2-k_{\bot}^2
-2 l_{\bot}k_{\bot}\cos \vartheta-m_3^2+i\varrho \\
{\cal P}_4&=&(k_{0}-q)^2-k_{\bot}^2-m_4^2+i\varrho \\
{\cal P}_5&=&k_{0}^2-k_{\bot}^2-m_5^2+i\varrho
\eeas
The angular integration is elementary. For the orthogonal space integration the
residue theorem is applied twice. The analytic result then is the following
two-fold integral representation (written for the scalar case in dimensionless
variables $x,y$).
\beas
\lefteqn{\hspace*{-2cm}\itg_{-\infty}^{\infty} d x\itg_{-\infty}
^{\infty} d y \,\,\frac{\log(w_3+w_2+w_5)
+ \{\mbox{similar terms} \}}{(w_1^2-w_2^2)(w_4^2-w_5^2)}} \\
w_1&=&\sqrt{x^2-\frac{m_1^2-i\varrho}{q^2}} \\
w_2&=&\sqrt{(x+1)^2-\frac{m_2^2-i\varrho}{q^2}} \\
w_3&=&\sqrt{(x+y)^2-\frac{m_3^2-i\varrho}{q^2}} \\
w_4&=&\sqrt{(y-1)^2-\frac{m_4^2-i\varrho}{q^2}} \\
w_5&=&\sqrt{y^2-\frac{m_5^2-i\varrho}{q^2}}
\eeas

\subsection{Numerical evaluation}

The two-fold integral representation can be directly taken for numerical
evaluation. We compared the results with available data.
The data coincide perfectly with results from Shimizu, Kato and
Fujimoto \cite{Fu2}. They were also verified in several asymptotic
limits \cite{Da2,Da3}.

\section{Future extensions}

The XLOOPS package is at present far from being saturated. The most important
extension will be the incorporation of two-loop three-point (and four-point)
functions. In addition a drawing routine for diagrams (generating Postscript
output) will also be included as well as the possibility of evaluating complete
processes and the option to renormalize the processes automatically.

\section*{Acknowledgements}

It is a great pleasure to thank L. Br\"ucher, A. Frink and D. Kreimer for the
friendly and stimulating atmosphere in which XLOOPS was developed. I also want
to thank J. G. K\"orner, U. Kilian, K. Schilcher and J. B. Tausk for many
discussions and encouragement. This work was supported by the Graduiertenkolleg
``Elementarteilchenphysik bei mittleren und hohen Energien" in Mainz. The
support by HUCAM, grant CHRX-CT94-0579, is also gratefully acknowledged.


\talk{Massive two-loop vertex functions}
{Alexander Frink\protect\footnote{email:
Alexander.Frink@Uni-Mainz.DE}}

Calculating massive two-loop vertex functions by splitting
integrations into parallel and orthogonal space components
has been demonstrated to give a convenient twofold integral
representation suitable for numerical evaluation.
This method is now extended to more topologies, including the
crossed topology, graphs with divergent subloops and infrared
divergent diagrams. The connection between this representation
and physical and anomalous thresholds is examined.

\section{The crossed vertex function}

The scalar crossed vertex function (Fig.~\ref{fig:tops}a) can be
calculated similarly to the planar vertex function
(Fig.~\ref{fig:tops}b) with the orthogonal/parallel space integration
technique presented in \cite{ckk}. The volume element is written as
$d^4l \, d^4k = \frac{1}{2} dl_0 \, dk_0 \, dl_1 \, dk_1 \, ds \, dt \,
d\alpha \, dz/\sqrt{1-z^2}$, where $l_0$, $l_1$, $k_0$ and $k_1$
are the components of the loop momenta parallel to the external
momenta, $s \equiv l_\perp^2$, $t \equiv k_\perp^2$ and $z$ is the
cosine of the angle between $\vec{l}_\perp$ and $\vec{k}_\perp$. 
The procedure can be divided into the following main steps:

\begin{figure}[ht]
\begin{center}
\epsfig{file=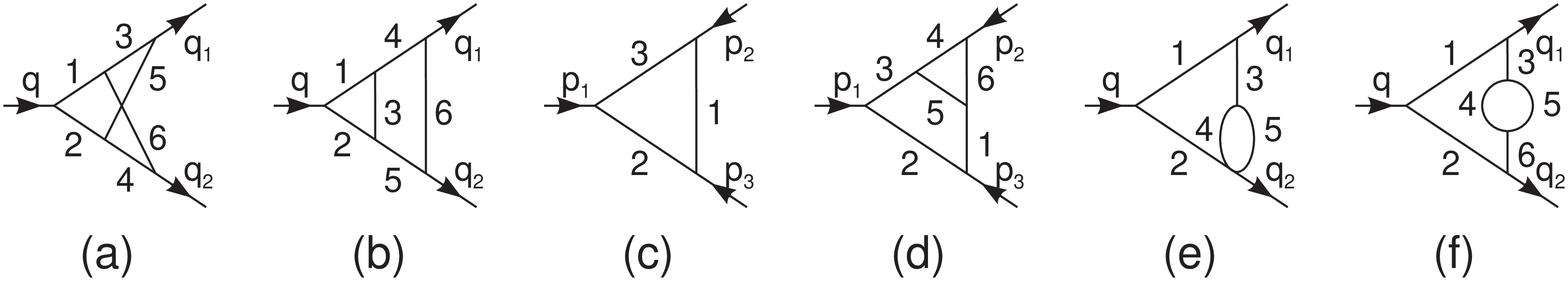,width=0.85\linewidth}
\end{center}
\caption{Topologies covered in this article}
\label{fig:tops}
\end{figure}

\begin{figure}[ht]
\begin{center}
\epsfig{file=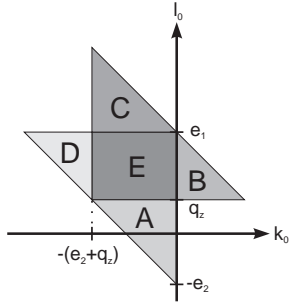,width=0.25\linewidth}
\end{center}
\caption{Effective integration region in the $(l_0,k_0)$ plane
         for the crossed vertex function}
\label{fig:region}
\end{figure}

\begin{itemize}
\item
linearization of propagators in $l_1$ and $k_1$ by shifting
$l_0 \to l_0 + l_1$, $k_0 \to k_0 + k_1$
\item
integration over orthogonal space angles ($\alpha$ and $z$)
\item 
integration over $l_1$ and $k_1$ with Cauchy's theorem, avoiding cuts
resulting from the $z$ integration
\item
integration over $s$ and $t$ with Euler's change of variables
\item
numerical evaluation of the remaining integrations over $l_0$ and
$k_0$ in a finite region
\end{itemize}
The main complication arises here as both loop momenta $l$ and $k$
have to flow through two propagators in common.
To perform the $z$ integration as in \cite{ckk}, we have to apply
a partial fraction decomposition in these two propagators,
in order that in the difference term the $z$ dependence drops out.
Each term of the partial fraction decomposition can be calculated
separately and similarly to the planar vertex function as in
\cite{ckk}, however the condition in which way to close the contour
for the $l_1$ and $k_1$ residue integrations is different for both
terms: $l_0+k_0-e_1 \lessgtr 0$ and $l_0+k_0+e_2 \lessgtr 0$ resp.,
where $e_1$ and $e_2$ are components of the external momenta.
As a consequence poles from the difference term which lie on the
real axis also have to be taken into account. These poles give rise
to several unbounded areas in the $(l_0,k_0)$ plane in addition to
triangles. Nevertheless these add up to zero outside the finite
region shown in Fig.~\ref{fig:region}.

After the residue integrations we have the intermediate result
\begin{eqnarray}
\label{eq:fourfold}
V_C & = & \sum_{j=1}^{10} \int\!\!\!\int\limits_{\!\! A_j} \! 
dl_0 \, dk_0 \int\limits_0^\infty \! ds \int\limits_0^\infty \! dt \,
C_j \,
\frac{1}{\tilde{a}_{1j} t + \tilde{b}_{1j} + \tilde{c}_{1j} c}
\frac{1}{\tilde{a}_{2j} t + \tilde{b}_{2j} + \tilde{c}_{2j} c}
\nonumber\\
& & \hspace{1cm}
\frac{1}{\tilde{a}_{3j} t + \tilde{b}_{3j} + \tilde{c}_{3j} c}
\frac{1}{\sqrt{(a_j t + b_j + c_j s)^2-4st}} \, ,
\end{eqnarray}
where $C_j$, $\tilde{a}_{kj}$, $\tilde{b}_{kj}$, $\tilde{c}_{kj}$,
$a_j$, $b_j$ and $c_j$ are rational functions of $l_0$ and $k_0$,
and $A_j$ is a subset of the area in Fig.~\ref{fig:region}.
Some of the $\tilde{a}_{kj}$ and $\tilde{c}_{kj}$ vanish.
Euler's change of variables for $s$ and $t$ reduces the problem
to integrals of the forms $\int \ln(x^2+rx+s)/(x^2+px+q) \, dx$
and $\int \arctan(x+a)/(x^2+px+q) \, dx$, which can be expressed
in terms of dilogarithms and Clausen functions.
Further details can be found in \cite{fkk}.

As for the planar vertex function a correlation between the various
coefficients in Eq.~(\ref{eq:fourfold})
and physical thresholds resulting in an 
imaginary part of the diagram can be found: if an $\tilde{a}_{kj}$
or $\tilde{c}_{kj}$ is zero, then the accompanying $\tilde{b}_{kj}$ 
corresponds to a two-particle cut. The coefficients $b_j$ 
correspond either to three-particle cuts or the
two-particle cut $q^2>(m_1+m_2)^2$ which is somewhat hidden due to
the partial fraction decomposition.

\section{Anomalous Thresholds}

\begin{figure}[ht]
\begin{center}
\setlength{\unitlength}{1cm}
\begin{picture}(11,3.8)
\put(0,0.3){\epsfig{file=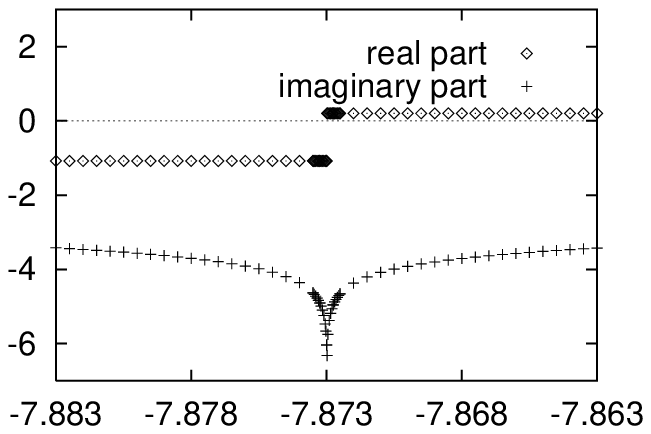,width=0.4\linewidth}}
\put(3,0){${\scriptstyle{\sf p_3^2}}$}
\put(5.5,0.3){\epsfig{file=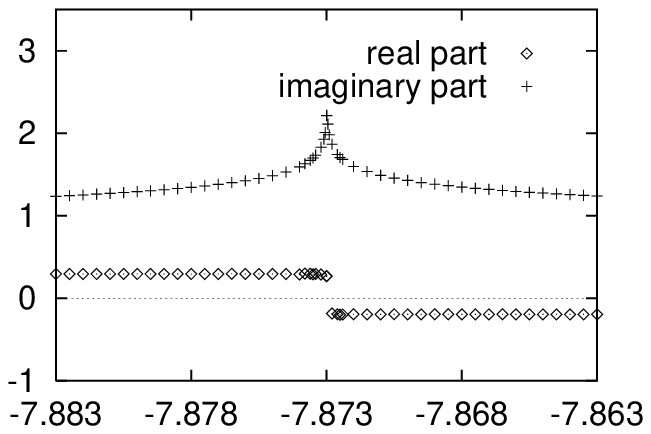,width=0.4\linewidth}}
\put(8.5,0){${\scriptstyle{\sf p_3^2}}$}
\end{picture}
\end{center}
\caption{Plot of real and imaginary parts of a vertex function in the
vicinity of an anomalous threshold, one-loop (left) and two-loop
(right)}
\label{fig:anom}
\end{figure}

The connection between the coefficients in our two-loop three-point
representation and physical thresholds has been pointed out in
\cite{ckk} for the planar and above for the crossed topology.
Now it will be demonstrated that anomalous thresholds are also
accounted for correctly in this representation.
As an example, let us consider the
one-loop three-point function in Fig.~\ref{fig:tops}c. By searching
for solutions of the Landau equations \cite{landau}, one can find a
singularity which corresponds to no physical threshold at
$1 + 2 \mu_1 \mu_2 \mu_3 - \mu_1^2 - \mu_2^2 - \mu_3^2 = 0$
where
$\mu_j = (m_1^2+m_2^2+m_3^2-m_j^2-p_j^2)m_j/(2 m_1 m_2 m_3)$.
If we arbitrarily choose all masses equal, $m_1=m_2=m_3=1$,
and $p_1^2=6$, $p_2^2=5$, these equations are fulfilled for
$p_3^2=-4-\sqrt{15} \approx -7.873$.
A plot of the real and imaginary part of this diagram for different
values of $p_3^2$ in the vicinity of this point is shown in
Fig.~\ref{fig:anom}(left). 

Let us now turn to the two-loop example Fig.~\ref{fig:tops}d.
The Landau equations give us a singularity for this graph at the
same values as above for arbitrary $m_4$, $m_5$ and $m_6$. We recall
the four-fold integral representation for the planar graph \cite{ckk}:
\begin{equation}
V_P = \sum\limits_{j=1}^4
      \int\!\!\!\int\limits_{\!\! T_j}\! dl_0 \, dk_0 \, C
      \int\limits_0^\infty \frac{dt}{(t+t_0)(t+t_0')}
      \int\limits_0^\infty \frac{ds}{s+s_0}
      \frac{1}{\sqrt{(at+b+cs)^2-4st}}
\end{equation} 
$t_0$ and $t_0'$ are quadratic functions in $k_0$. As was pointed out
in \cite{ckk}, real roots of $t_0$ and $t_0'$ inside the integration
triangle correspond to physical thresholds $p_1^2 > (m_2+m_3)^2$
and $p_2^2 > (m_3+m_1)^2$ respectively. These conditions are fulfilled
here. Let us denote the roots of $t_0$ and $t_0'$ as $k_0^{(1,2)}$
and ${k_0^{(1,2)}}{{}'}$ resp. Then it can be shown that below
the anomalous threshold $k_0^{(2)} < k_0^{(1)}{}'$ and above
$k_0^{(2)} > k_0^{(1)}{}'$.
However, for the numerical integration over $l_0$ and $k_0$
(after partial fraction decomposition) the program only has to
make a distinction between $t_0^{(\prime)}>0$ (real part only)
and $t_0^{(\prime)}<0$ (additional imaginary part).
Merely numerical stability gets slightly worse in the direct
vicinity of the threshold. Fig.~\ref{fig:anom}(right) shows the
expected plot of real and imaginary parts.

\section{Divergent diagrams}

The method for calculating massive two-loop vertex functions
by splitting integrations into parallel and orthogonal space
components works without modifications only for convergent integrals.
The complication stems from the non-elementary integration over the
cosine of the angle between $\vec{l}_\perp$ and $\vec{k}_\perp$, $z$,
in $D=4-2\varepsilon$ dimensions
\begin{equation}
\int\limits_{-1}^{1} \frac{1}{\sqrt{1-z^2}}
                     \frac{1}{A + B z + i \eta} \, dz
\longrightarrow
\int\limits_{-1}^{1} (1-z^2)^{\frac{D-5}{2}} 
                     \frac{1}{A + B z + i \eta} \, dz
\, ,
\end{equation}
which can only be expressed in terms of hypergeometric functions
instead of a simple square root.

In order to avoid this, one has to subtract a simpler function with
the same structure of divergence. The difference is finite and can be 
calculated with the parallel/orthogonal space technique in four
dimensions, whereas the subtracted function should be calculable
in $D$ dimensions. A first example for an UV divergent scalar vertex
function at zero momentum transfer has been given in \cite{andrzej}.
This will now be extended to other genuine vertex topologies
as well as infrared divergent diagrams.

\subsection{Infrared divergences}

As an example let us consider the planar graph depicted in
Fig.~\ref{fig:tops}b with $q_1^2=m_4^2$, $q_2^2=m_5^2$ and
$m_6 \to 0$. We choose the loop momentum $l$ to flow through
$P_1 \dots P_3$ and $k$ through $P_4 \dots P_6$. This graph exhibits
an infrared divergence for $k \to 0$. By subtracting the same
diagram, but with $P_3$ replaced with $\tilde{P}_3 = P_3 \big|_{k=0}$,
power-counting for small $k$ is improved by one unit, so the
difference is now finite and can be calculated in four dimensions:
\begin{eqnarray}
V_{IR} & = & \int\! d^4l \, d^4k \, \frac{1}{P_1 P_2}
\left(\frac{1}{P_3}-\frac{1}{\tilde{P}_3}\right) \frac{1}{P_4 P_5 P_6}
\nonumber\\
& & + \int\! d^Dl \, \frac{1}{P_1 P_2 \tilde{P}_3}
      \int\! d^Dk \, \frac{1}{P_4 P_5 P_6} + {\cal O}(\varepsilon)
\, .
\end{eqnarray}
This involves the same main steps as for the crossed vertex function 
(see above). Since infrared divergences are endpoint singularities,
all steps can be applied individually to both parts of the integrand,
and the infrared divergence will not show until the last numerical
integration over $k_0$ near $k_0 \to 0$. But there the subtraction
guarantees a well-defined limit.

Let us now demonstrate the integration steps for the subtracted term.
The $z$ integration is trivial, since none of the propagators depends
on $z$. Therefore there is no cut in the integrand which has to be
avoided during the residue integrations. We have the freedom to 
close the contours analogously to the original term \cite{ckk}, i.e.
for $l_0+k_0>0$ in the upper and for $l_0+k_0<0$ in the lower
half plane. Then the integrations over $l_1$ and $k_1$ give conditions
of the form $l_0 \lessgtr x$ and $k_0 \lessgtr y$ for the
individual propagators to contribute, which results in triangular
regions listed in Table~\ref{tab:irtriangles}. The overlap of these
triangles coincides with the overlap of the triangles from \cite{ckk}.

\begin{table}[htbp]
\caption{{\label{tab:irtriangles}}Non-vanishing triangles
                                  from the subtracted term}
\begin{center}
\begin{tabular}{c c c c}\small\\
\hline
propagators & \multicolumn{3}{c}{conditions} \\
\hline
$(P_1,P_5)$ & $l_0+k_0 < 0 $ & $l_0+e_1-q_z > 0$ & $k_0+e_2+q_z > 0$ \\
$(P_1,P_6)$ & $l_0+k_0 < 0 $ & $l_0+e_1-q_z > 0$ & $k_0 > 0$ \\
$(P_2,P_4)$ & $l_0+k_0 > 0 $ & $l_0-e_2-q_z < 0$ & $k_0-e_1+q_z < 0$ \\
$(P_2,P_6)$ & $l_0+k_0 > 0 $ & $l_0-e_2-q_z < 0$ & $k_0 < 0$ \\
$(P_3,P_4)$ & $l_0+k_0 > 0 $ & $l_0 < 0$ & $k_0-e_1+q_z < 0$ \\
$(P_3,P_5)$ & $l_0+k_0 < 0 $ & $l_0 > 0$ & $k_0+e_2+q_z > 0$ \\
\hline\\
\end{tabular}
\end{center}
\end{table}

The analytic evaluation of the subtracted part involves calculating
one-loop three-point functions up to ${\cal O}(\varepsilon)$
\cite{nierste}. This subtraction procedure is also applicable for
small but non-zero $m_6$ in order to improve numerical stability and
to extract the leading logarithm.

\subsection{Graphs with divergent subloops}

Ultraviolet divergent scalar diagrams as in Fig.~\ref{fig:tops}e
can also be calculated by subtracting an appropriate quantity with
the same structure of divergence. This can be done by setting the
masses in the inner subloop to zero. Let $k$ flow through the
divergent subloop and $l$ in the outer loop. Defining
$\tilde{P}_4 = P_4 \big|_{m_4=0}$ and
$\tilde{P}_5 = P_5 \big|_{m_5=0}$, we have
\begin{eqnarray}
V_{UV} & = & \int \! d^4l \, d^4k \frac{1}{P_1 P_2 P_3}
\left(\frac{1}{P_4 P_5}-\frac{1}{\tilde{P}_4 \tilde{P}_5}\right)
\nonumber\\
& & + \int\! d^Dl \int\! d^Dk
      \frac{1}{P_1 P_2 P_3 \tilde{P}_4 \tilde{P}_5}
    + {\cal O}(\varepsilon)
\, .
\end{eqnarray}
Performing the $z$, $l_1$ and $k_1$ integrations as above, we obtain
the intermediate result
\begin{eqnarray}
V_{UV, {\rm finite}} & = & \sum_{j=1}^2 
\int\!\!\!\int \! dl_0 \, dk_0 \int\limits_0^\infty \! ds 
\int\limits_0^\infty \! dt 
\, C \, \frac{1}{(s+s_0)(s+s_0')}
\nonumber\\
& & \hspace{1cm}
\left( 
\frac{1}{\sqrt{(at+b+cs)^2-4st}} -
\frac{1}{\sqrt{(at+\tilde{b}+cs)^2-4st}}
\right)
\, .
\end{eqnarray}
This can be readily expressed as a twofold integral representation 
over dilogarithms. The $l_0$ and $k_0$ integrations extend again
over triangles. Note that the boundaries of the triangles depend
solely on the external momenta and not on the internal masses.

The calculation of the subtracted part is done by first integrating
the massless subloop in $k$, giving a result proportional to
$(l^2)^{-\varepsilon}$. The evaluation of the one-loop three-point
function with non-integer powers of propagators can be done at least
numerically.

The topology in Fig.~\ref{fig:tops}f can be reduced to the topology in
Fig.~\ref{fig:tops}e by partial fraction decomposition in the
propagators $P_3$ and $P_6$. Since the integration region in the
$(l_0,k_0)$ plane is equal for both terms, they can be added again
before integrating over $l_0$ and $k_0$. The case $m_3=m_6$ is
handled by differentiating $V_{UV}$ with respect to $m_3$.

\section{Outlook}

Current work is done in the direct calculation of tensor integrals.
A general method using subtraction terms similar to above has been
proposed by D.~Kreimer \cite{dirktensor}.

\section*{Acknowledgements}

I would like to thank D.~Kreimer, K.~Schilcher, J.B.~Tausk,
J.~Franzkowski and L.~Br\"{u}cher for helpful discussions and
valuable inputs on this subject. This work is supported by
``Graduiertenkolleg Elementarteilchenphysik bei mittleren und
hohen Energien'' of Univ\@. Mainz.

\end{document}